\documentclass[aps,prd,twocolumn,superscriptaddress,groupedaddress,nofootinbib]{revtex4}  

\usepackage{amsmath,amssymb,bm,color,comment,dcolumn,mathrsfs,tabularx,graphicx}
\usepackage[hidelinks=true]{hyperref}
\hypersetup{colorlinks=true,linkcolor=blue,citecolor=blue,urlcolor=blue}

\def\l{\ell}

\def\ave#1{\langle #1 \rangle}
\def\vs #1{\vspace{#1\baselineskip}}
\newcommand\E{\hspace{0.1em}{\rm e}}

\newcommand{\mC}[1]{\mathcal{#1}}   
\newcommand{\dd}[2]{\,{\rm d}^{#1} #2 \,} 
\newcommand{\al}[1]{\ifdefined\compile \begin{align} #1 \end{align}\fi}
\newcommand{\INT}[5]{{\int_{#4}^{#5} \!\! \dd{#1}{#2} \,}}
\newcommand{\bc}{\begin{center}}
\newcommand{\ec}{\end{center}}

\def\bl{\bm{\ell}}

\def\bk{\bm{k}}

\def\grad{\phi}

\newcommand{\Wjm}[6]{\begin{pmatrix} #1 & #2 & #3 \\ #4 & #5 & #6 \end{pmatrix}}
\newcommand{\compile}{}

\date{\today}

\begin{document}

\ifdefined\compile 

\title{CMB Lensing Bispectrum from Nonlinear Growth of the Large Scale Structure}

\author{Toshiya Namikawa}
\affiliation{Department of Physics, Stanford University, Stanford, California 94305, USA}
\affiliation{Kavli Institute for Particle Astrophysics and Cosmology, SLAC National Accelerator
Laboratory, Menlo Park, California 94025, USA}

\date{\today}


\begin{abstract}
We discuss detectability of the nonlinear growth of the large-scale structure in the cosmic microwave 
background (CMB) lensing. The lensing signals involved in the CMB fluctuations have been measured 
from multiple CMB experiments, such as Atacama Cosmology Telescope (ACT), Planck, POLARBEAR, and 
South Pole Telescope (SPT). The reconstructed CMB lensing signals are useful to constrain cosmology 
via their angular power spectrum, while detectability and cosmological application of their bispectrum 
induced by the nonlinear evolution are not well studied. Extending the analytic estimate of the galaxy 
lensing bispectrum presented by Takada and Jain (2004) to the CMB case, we show that even near term 
CMB experiments such as Advanced ACT, Simons Array and SPT3G could detect the CMB lensing bispectrum 
induced by the nonlinear growth of the large-scale structure. In the case of the CMB Stage-IV, we find 
that the lensing bispectrum is detectable at $\agt 50\sigma$ statistical significance. This precisely 
measured lensing bispectrum has rich cosmological information, and could be used to constrain 
cosmology, e.g., the sum of the neutrino masses and the dark-energy properties.  
\end{abstract}

\maketitle

\fi 


\section{Introduction} \label{intro}

The cosmic microwave background (CMB) temperature and polarization anisotropies are distorted by 
gravitational weak lensing on arcminute scales. From the past several years, the lensing signals 
involved in observed CMB anisotropies have been detected from multiple CMB experiments, such as 
Atacama Cosmology Telescope (ACT) \cite{Das:2011,Das:2014}, Planck \cite{P13lens,P15lens}, 
POLARBEAR \cite{PBCMB}, and South Pole Telescope (SPT) \cite{vanEngelen:2012va,Story:2014hni}. 
The reconstructed lensing signals have also been used to cross-correlate with galaxies/quasars
(e.g., \cite{Smith07,Hirata:2008cb,Bleem:2012gm,Sherwin:2012mr,Geach:2013zwa}) and cosmic infrared 
background (e.g., \cite{P13CIB,Holder:2013,Hanson:2013daa}).

Precisely measured lensing power spectra are already being used to constrain, e.g., the dark energy 
\cite{Sherwin:2011gv,vanEngelen:2012va,P13lens}, the sum of the neutrino masses \cite{Battye:2013},
the non-Gaussianity of the primordial density perturbation \cite{Giannantonio:2013}, a specific model of 
the cosmic string network \cite{Namikawa:2013a}, and a model of the dark-matter dark-energy interaction 
\cite{Wilkinson:2013kia}. Future CMB experiments are expected to quantify the sum of the neutrino masses 
(e.g., \cite{Namikawa:2010re,Abazajian:2013oma,Wu:2014hta,Allison:2015} and references therein),
and provide even tighter constraints on the dark energy, primordial non-Gaussianity, cosmic strings, 
and other fundamental physics. The ongoing and future CMB experiments such as Advanced ACT \cite{AdvACT}, 
Simons Array \cite{SimonsArray}, SPT3G \cite{SPT3G}, and Stage-IV \cite{Abazajian:2013oma} 
will significantly improve their sensitivity to CMB polarization at few arcminute scales, 
and realize very precise measurements of CMB lensing signals. 

So far, multiple studies have focused on cosmological applications of reconstructed lensing signals 
using their angular power spectrum (two-point correlation). On the other hand, the possibility of 
detecting the bispectrum (three-point correlation) of the lensing signals has not been studied. 
In general, Gaussian fluctuations $\delta$ do not produce bispectrum, but fluctuations including the 
non-Gaussian part ($\delta+\delta^2+\cdots$) give rise to the nonzero bispectrum, such as 
$\ave{\delta\delta\delta^2}$ or $\ave{\delta^2\delta^2\delta^2}$. Since the nonlinear growth of 
the large-scale structure (LSS) produces the bispectrum of the lensing signals, measurements of 
the lensing bispectrum provide additional cosmological information involved in the nonlinear clustering. 
However, compared to other lensing measurements such as galaxy weak lensing, the nonlinear evolution of 
the density fluctuations does not significantly affect the lensing of the CMB. The effect of 
the nonlinear growth on the lensing power spectrum is only important at $\l\agt 1000$ \cite{Lewis:2006}.
The weakness of the nonlinear effect comes from the fact that the gravitational potential which causes 
the CMB lensing is at relatively high-redshifts ($z\sim\mC{O}(1)$) \cite{Lewis:2006} where 
the gravitational potential would be described with linear perturbation theory at even small scales.
In addition, the lensing potential is a line-of-sight integral over the dark matter structures, and is 
a sum of the gravitational potential at different redshifts. This summation along the line-of-sight 
Gaussianizes the lensing potential by the central limit theorem \cite{Hirata:2003ka}, and 
the lensing bispectrum becomes small.

In past and ongoing CMB experiments which cannot extract small-scale lensing signals, the nonlinear 
effect does not significantly bias the cosmological parameter estimations. However, future CMB 
experiments such as CMB Stage-IV (S4) will realize very deep polarization measurement up to few 
arcminute scales, and precision of the lensing reconstruction will be improved significantly. In 
upcoming and future CMB experiments, the nonlinear growth of the LSS will be no longer negligible in 
the CMB analysis, and cause bias in several ways. Challinor and Lewis \cite{Challinor:2005jy} showed 
that B-mode polarization generated by lensing has $\sim 10$\% of the nonlinear contributions even at 
large scale $\l<100$. Boehm \textit{et al}. \cite{Boehm:2016} discuss how the presence of 
the lensing bispectrum biases the lensing reconstruction in S4, finding that there is a non-negligible 
contribution from the bispectrum in the power spectrum estimate. 

In this paper, we consider for the first time the nonlinear effect as a cosmological signal rather 
than a source of the bias, and discuss the detectability and cosmological application of 
the lensing bispectrum in CMB Stage-III (S3) and S4 experiments. We extend 
Takada and Jain (2004) \cite{TJ04} (hereafter TJ04) to the case with the CMB lensing to compute 
the lensing bispectrum and its signal to noise. We then discuss the cosmological application of 
the lensing bispectrum. 

This paper is organized as follows. In Sec.~\ref{bisp}, we summarize our analytic method to compute 
the lensing bispectrum. In Sec.~\ref{forecast}, we show the expected signal to noise of 
the lensing bispectrum from near term and future CMB experiments. We also show the impact of 
the inclusion of the lensing bispectrum in the cosmological parameter constraints. 
Section \ref{summary} is devoted to a summary. 

Throughout this paper, for the fiducial model, we assume the spatially flat $\Lambda$CDM cosmology 
with two massive ($m_{\nu,1}=0.05$ eV and $m_{\nu,2}=0.01$ eV mass eigenstates) and one massless neutrinos. 
We use a set of cosmological parameters consistent with the latest Planck results \cite{P15main}; 
the baryon and matter density, $\Omega_{\rm b}h^2=0.0223$ and $\Omega_{\rm m}h^2=0.119$, 
the dark-energy density $\Omega_\Lambda=0.689$, the amplitude of the primordial scalar 
power spectrum, $A_{\rm s}=2.13\times 10^{-9}$, and its spectral index at $k=0.05$ Mpc$^{-1}$, $n_{\rm s}=0.965$, 
and the reionization optical depth $\tau=0.0630$.
We employ {\tt CAMB} \cite{Lewis:1999bs} to compute the linear matter power spectrum at $z=0$ and lensing power spectrum. 
The linear matter power spectrum at $z=0$ is then used to evaluate those at each redshift $z$. 
We employ the fitting formula given by Refs.~\cite{Smith:2003,Takahashi:2012} 
to compute the nonlinear matter power spectrum from the linear matter power spectrum.

\section{CMB Lensing Bispectrum} \label{bisp}

Here we summarize the bispectrum of the CMB lensing signals. Our method to compute the CMB lensing 
bispectrum is based on TJ04. 

The distortion effect of lensing on the primary CMB anisotropies is expressed by a remapping 
(e.g. \cite{Lewis:2006}). This introduces statistical anisotropy into the observed CMB, in the form of 
a correlation between the CMB anisotropies and their gradient \cite{Hu:2001kj}. With a large number of 
observed CMB modes, this correlation is used to estimate the lensing signals involved in observed CMB 
anisotropies \cite{Zaldarriaga:1998te,Hu:2001kj,Hirata:2002jy,Hanson:2009gu}. The power spectrum of 
the lensing signals is in turn studied by taking the power spectrum of these lensing estimates 
\cite{Hu:2001kj,Namikawa:2012pe}.

The reconstructed lensing signals are also used to study other statistics such as the bispectrum. 
The bispectrum of the CMB lensing-mass (convergence) fields is defined as
\al{
	\ave{\kappa_{\l_1m_1}\kappa_{\l_2m_2}\kappa_{\l_3m_3}} 
		= \Wjm{\l_1}{\l_2}{\l_3}{m_1}{m_2}{m_3}B^\kappa_{\l_1\l_2\l_3} \,,
}
where $\kappa_{\l m}$ is the spherical harmonic coefficients of the lensing-mass fields, and 
$\ave{\cdots}$ denotes the ensemble average. The multipoles satisfy the triangle condition, 
$|\l_i-\l_j|\leq\l_k\leq\l_i+\l_j$. 

Note that, in the flat sky, with the Dirac delta function $\delta_{\rm D}$, the bispectrum is given by 
\al{
	\ave{\kappa_{\bl_1}\kappa_{\bl_2}\kappa_{\bl_3}} 
		= (2\pi)^2\delta_{\rm D}(\bl_1+\bl_2+\bl_3)B^\kappa(\bl_1,\bl_2,\bl_3) \,,
}
where $\kappa_{\bl}$ is the two-dimensional Fourier mode of the lensing-mass fields. 
The full-sky bispectrum is then approximately related to the flat-sky bispectrum as \cite{TJ04}
\al{
	&B^\kappa_{\l_1\l_2\l_3} \simeq \Wjm{\l_1}{\l_2}{\l_3}{0}{0}{0}
		\notag \\
	&\qquad \times \sqrt{\frac{(2\l_1+1)(2\l_2+1)(2\l_3+1)}{4\pi}}B^\kappa(\bl_1,\bl_2,\bl_3)
	\,. \label{Eq:full-bisp}
}
Denoting $L=(\l_1+\l_2+\l_3)/2$, an approximate form of the Wigner 3j symbol is given by \cite{TJ04}
\al{
	&\Wjm{\l_1}{\l_2}{\l_3}{0}{0}{0} \simeq (-1)^L\sqrt{\frac{\E}{2\pi}}(L+1)^{-1/4}
		\notag \\
	&\qquad \times \prod_{i=1}^3(L-\l_i+1)^{-1/4}\left(\frac{L-\l_i+1/2}{L-\l_i+1}\right)^{L-\l_i+1/4}
	\,. \label{Eq:wig-approx}
}
In our calculation, we evaluate the full-sky bispectrum from the flat-sky bispectrum given by 
Eqs.~\eqref{Eq:full-bisp} and \eqref{Eq:wig-approx}. 

The lensing-mass field is the line-of-sight integral over the gravitational potential of the LSS, 
and is directly related to the matter inhomogeneities. 
The flat-sky lensing bispectrum is expressed in terms of the three-dimensional matter bispectrum as
\al{
	B^\kappa(\bl_1,\bl_2,\bl_3) = \INT{}{\chi}{}{0}{\chi_*} 
		\frac{W^3(\chi)}{\chi^4}B^{\rm m}(\bk_1,\bk_2,\bk_3,\chi)
	\,, \label{Eq:flat-bisp}
}
where $\bk_i=\bl_i/\chi$, and the lensing kernel is given by
\al{
	W(\chi) = \frac{3\Omega_{\rm m,0}H_0^2}{2a(\chi)}\frac{\chi(\chi_*-\chi)}{\chi_*}  \,. 
}
The quantities, $\chi$, $\Omega_{\rm m,0}$, $a$, and $H_0$, denote the comoving distance, matter 
energy density, scale factor, and current expansion rate of the Universe. $\chi^*$ is the comoving 
distance to the last scattering surface of CMB photons. Compared to the galaxy lensing, 
the above lensing kernel is sensitive to the structure at relatively high redshifts [$z\sim\mC{O}(1)$] \cite{Lewis:2006}. 
In addition, since $\l=k\chi$, for a given multipole $\l$, CMB lensing signals pick up the density 
fluctuations at larger scales compared to the galaxy lensing. The lensing signals are therefore less 
sensitive to the nonlinear growth of the LSS at the late time of the Universe. 

The matter bispectrum is induced by the nonlinear clustering in the matter density fluctuations. 
The thorough calculation of the nonlinear clustering, however, requires an expensive numerical simulations. 
To clarify the feasibility of the lensing bispectrum in cosmology, TJ04 computed the matter bispectrum 
using a fitting formula for the matter bispectrum. We follow TJ04 and use the best available fitting 
formula for the matter bispectrum which is recently developed by Ref.~\cite{Gil-Marin:2012} and is 
an extension of the formula in Ref.~\cite{Scoccimarro:2001}. In this formula, the matter bispectrum is given by
\al{
	B^{\rm m}(\bk_1,\bk_2,\bk_3,\chi) &= 2F_2(\bk_1,\bk_2,z)P_{\rm m}(k_1,z)P_{\rm m}(k_2,z) 
	\notag \\
		&+ 2\, {\rm perms.}
	\,, \label{Eq:bisp-pk}
}
where $P_{\rm m}(k,z)$ is the nonlinear matter power spectrum and, with 
$\bk_1\cdot\bk_2=k_1k_2\cos\theta$, the ``effective'' F2 kernel is defined as \cite{Scoccimarro:2001}
\al{
	&F_2(\bk_1,\bk_2,z) = \frac{5}{7}a(k_1,z)a(k_2,z) \notag \\
	&\qquad + \frac{k_1^2+k_2^2}{2k_1k_2}b(k_1,z)b(k_2,z)\cos\theta \notag \\
	&\qquad + \frac{2}{7}c(k_1,z)c(k_2,z)\cos^2\theta \,. \label{Eq:F2}
}
The factors $a(k,z)$, $b(k,z)$ and $c(k,z)$ are defined by \cite{Scoccimarro:2001}
\al{
	a(k,z) &= \frac{1+\sigma_8^{a_6}(z)\sqrt{0.7Q(n_{\rm eff})}(qa_1)^{n_{\rm eff}+a_2}}{1+(qa_1)^{n_{\rm eff}+a_2}} \\
	b(k,z) &= \frac{1+0.2a_3(n_{\rm eff}+3)(qa_7)^{n_{\rm eff}+3+a_8}}{1+(qa_7)^{n_{\rm eff}+3.5+a_8}} \\
	c(k,z) &= \frac{1+[4.5a_4/(1.5+(n_{\rm eff}+3)^4)](qa_5)^{n_{\rm eff}+3+a_9}}{1+(qa_5)^{n_{\rm eff}+3.5+a_9}}
	\,, 
}
with $Q(x)=(4-2^x)/(1+2^{x+1})$. The quantity, $\sigma_8(z)$ is the variance of the matter density 
fluctuations smoothed by $8$Mpc $h^{-1}$ at redshift $z$. 
$n_{\rm eff}\equiv d\ln P_{\rm m}^{\rm lin}(k)/dk$ is the effective spectral index of the power spectrum, 
with $P_{\rm m}^{\rm lin}(k)$ being the linear matter power spectrum. We define $q=k/k_{\rm NL}$ with 
the nonlinear scale $k_{\rm NL}$ satisfying $4\pi k^3_{\rm NL}P_{\rm m}^{\rm lin}(k_{\rm NL})=1$. 
The coefficients of the fitting formula $a_i$ ($i=1,2,\dots,9$) are obtained by fitting to 
the simulations in Ref.~\cite{Gil-Marin:2012}, and their best fit parameters are
$a_1 = 0.484$, $a_2 = 3.740$, $a_3 = -0.849$, $a_4 = 0.392$, $a_5 = 1.013$, $a_6 = -0.575$, 
$a_7 = 0.128$, $a_8 = -0.722$, and $a_9 = -0.926$. 

At $k\ll k_{\rm NL}$ ($q\ll 1$), the effective F2 kernel recovers the second order (tree-level) 
prediction of Eulerian perturbation theory in an Einstein de-Sitter universe. 
The tree-level bispectrum is computed with $P_{\rm m}^{\rm lin}$ in Eq.~\eqref{Eq:bisp-pk}, and 
$a=b=c=1$ in Eq.~\eqref{Eq:F2}. 

\begin{figure}[t]
\bc
\ifdefined\compile
\includegraphics[width=8.5cm,height=6cm,clip]{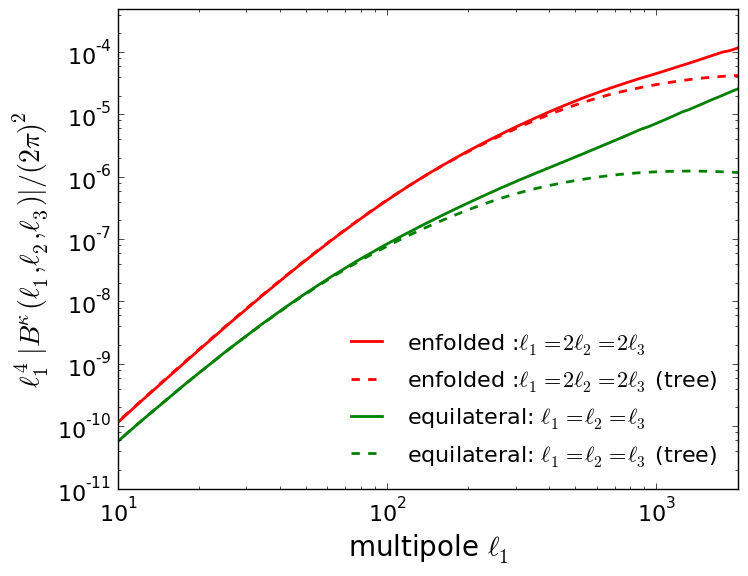} 
\fi
\caption{
The lensing bispectra with the enfolded ($\l_1=2\l_2=2\l_3$; red) and equilateral ($\l_1=\l_2=\l_3$; green) 
triangular configurations in the multipole space. The dashed lines (denoted by ``tree'') show 
the bispectrum from the tree-level prediction.
}
\label{fig:blll}
\ec
\end{figure}

\begin{figure}[t]
\bc
\ifdefined\compile
\includegraphics[width=8.5cm,height=6cm,clip]{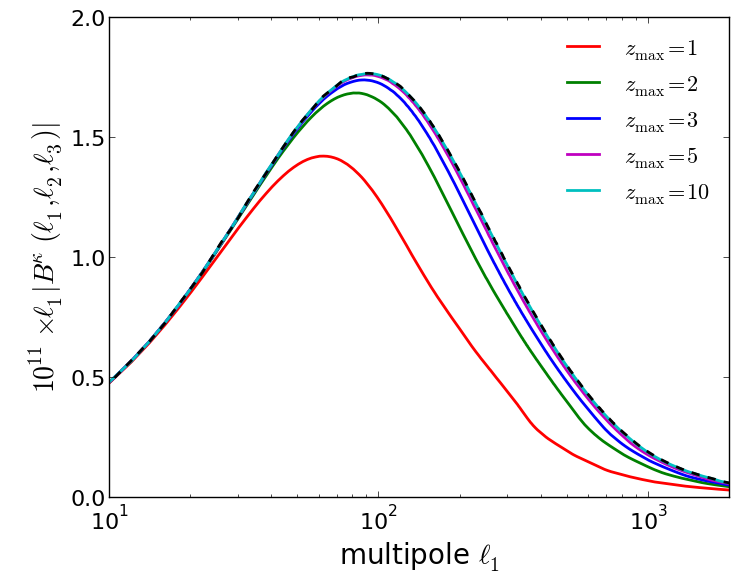} 
\fi
\caption{
The contribution to the enfolded ($\l_1=2\l_2=2\l_3$) lensing bispectrum from the gravitational 
potential of the LSS between $z=0$ and $z_{\rm max}$. The black dashed line shows the full lensing 
bispectrum. The maximum redshift is varied from $z_{\rm max}=1$ to $10$. For illustrative purposes, 
we multiply $10^{11}\l_1$ to the lensing bispectrum.
}
\label{fig:blll-z}
\ec
\end{figure}

Figure \ref{fig:blll} shows the CMB lensing bispectrum computed from Eq.~\eqref{Eq:flat-bisp}. We show 
the following two cases: the enfolded ($\l_1=2\l_2=2\l_3$) and, for comparison with TJ04, the 
equilateral ($\l_1=\l_2=\l_3$) triangular configurations. We also plot the bispectra from the tree 
level prediction. As we discussed later, the enfolded bispectrum provides the most part of the signal 
to noise. The tree-level bispectrum starts to deviate from the full nonlinear bispectrum at 
$\l\sim\mC{O}(100)$. The enfolded bispectrum has no significant corrections from the full nonlinear 
effect even at $\l\sim1000$. On the other hand, in the case of the galaxy weak lensing, the full 
nonlinear contribution is important even at large scales $\l\sim\mC{O}(10)$ \cite{TJ04}, and the 
tree-level contribution is several orders of magnitude smaller than the full nonlinear contribution at $\l=1000$. 
This fact indicates that, compared to the galaxy lensing bispectrum, 
the CMB lensing bispectrum is much less sensitive to the nonlinear clustering beyond the tree level. 
Note that the amplitude of the enfolded bispectrum is larger than that of the equilateral bispectrum. 
This is because the matter bispectrum at smaller scales (equivalent to low redshifts for a given multipole) 
which produces most of the lensing signals has large amplitudes in the enfolded configuration 
\cite{Jeong:2009,Schmittfull:2012}. 

Figure \ref{fig:blll-z} shows the contributions from $z=0$ to $z_{\rm max}$ to the enfolded lensing 
bispectrum. The lensing bispectrum is mostly generated from lower redshifts $z\leq 3$. 
The contributions from higher redshifts $z\geq 5$ are negligible in the lensing bispectrum. We also 
check that the equilateral case has also similar $z_{\rm max}$ dependence.

\section{Cosmological Forecasts} \label{forecast}

We next show the expected signal to noise of the lensing bispectrum from the near future CMB experiments.

\subsection{Detectability of the lensing bispectrum} \label{result:snr}

\begin{table}
\bc
\caption{
Experimental specifications for CMB experiments considered in this paper: the noise level in 
the polarization map ($\Delta_{\rm P}$) in unit of $\mu$K-arcmin, the beam size of FWHM ($\theta$) 
in unit of arcmin, and fractional sky coverage $f_{\rm sky}$.
}
\label{Table:cmb} \vs{0.5}
\begin{tabular}{lccc} \hline 
 & $\Delta_{\rm P}$ [$\mu$K-arcmin] & $\theta$ [arcmin] & $f_{\rm sky}$ \\ \hline 
S3-wide & 6 & 1 & 0.5  \\ 
S3-deep & 3 & 1 & 0.05 \\ 
S4      & 1 & 3 & 0.5  \\ \hline
\end{tabular}
\ec
\end{table}

\begin{figure}[t]
\bc
\ifdefined\compile
\includegraphics[width=8.5cm,height=6cm,clip]{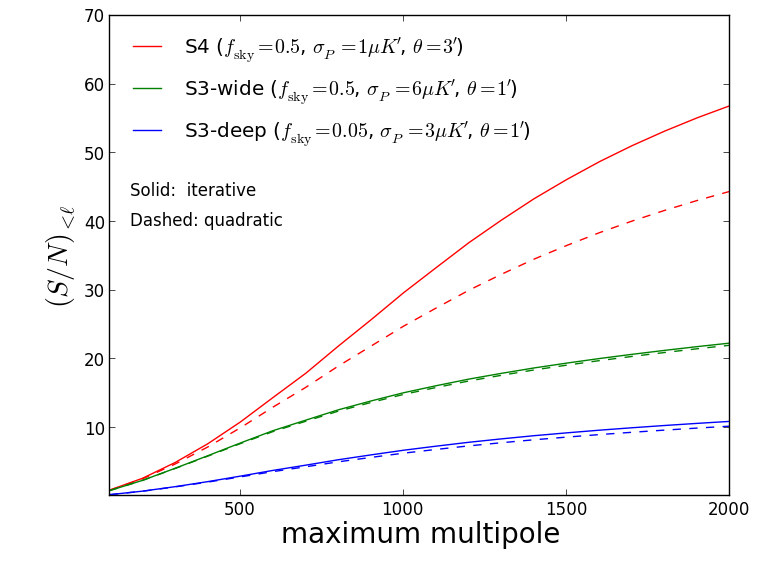} 
\fi
\caption{
Expected cumulative signal to noise ratios of the CMB lensing bispectrum with the S3-wide (such as 
Advanced ACT and Simons Array), the S3-deep (such as SPT3G) and the CMB Stage-IV. The dashed lines 
show the cases with the noise spectra computed from the quadratic estimator of Ref.~\cite{Hu:2001kj}. 
}
\label{fig:snr}
\ec
\end{figure}

We estimate the detectability of the lensing bispectrum as \cite{TJ04}
\al{
	\left(\frac{S}{N}\right)^2_{\leq\l} 
		&= f_{\rm sky} \sum^\l_{\l_1\leq\l_2\leq\l_3} 
		\frac{(B^\kappa_{\l_1\l_2\l_3})^2}{\Delta_{\l_1\l_2\l_3}\mC{C}_{\l_1}\mC{C}_{\l_2}\mC{C}_{\l_3}}
	\,, \label{Eq:snr}
}
where $\mC{C}_\l$ is the sum of the signal and noise power spectrum in the CMB lensing reconstruction. 
We define $\Delta_{\l_1\l_2\l_3}=1$ if all $\l_i$ are different, $\Delta_{\l_1\l_2\l_3}=2$ if two of 
$\l_i$ are equal, and $\Delta_{\l_1\l_2\l_3}=6$ if all of $\l_i$ are equal, respectively. The Gaussian 
covariance of the lensing bispectrum is assumed in the above equation. The noise power spectrum of 
the lensing signals, $N^{\grad}_\l$, is computed based on the iterative estimator developed in 
Refs.~\cite{Hirata:2003ka,Smith:2010gu}. 

To compute the noise power spectrum of the lensing signals, $N^{\grad}_\l$, in Eq.~\eqref{Eq:snr}, we 
assume a white noise with a Gaussian beam. We employ the formula of the noise power spectrum in 
Ref.~\cite{Knox:1999} which is characterized by the following two parameters; the noise level 
$\Delta_{\rm P}$ in unit of $\mu$K-arcmin, and beam size of FWHM $\theta$ in unit of arcmin. 
The summary of the experimental specifications are given in Table.~\ref{Table:cmb}. We denote 
``S3-wide'' as a wide Stage-III class experiment such as Advanced ACT and Simons Array, and ``S3-deep'' 
as a deep Stage-III class experiment such as SPT3G. We choose $\Delta_{\rm P}=6\mu$K-arcmin and 
$\theta=1$ arcmin for the S3-wide, $\Delta_{\rm P}=3\mu$K-arcmin and $\theta=1$ arcmin for the S3-deep, 
and $\Delta_{\rm P}=1\mu$K-arcmin and $\theta=3$ arcmin or the S4. For the S3-wide and S4 experiments, 
we assume the fractional survey area as $f_{\rm sky}=0.5$, while we choose $f_{\rm sky}=0.05$ for S3-deep. 
The CMB multipoles up to $\l=4000$ are used to estimate the noise power spectrum. 

Figure \ref{fig:snr} shows the signal-to-noise ratio of the S3-wide, S3-deep and S4 experiments. Even 
the Stage-III experiments would detect the CMB lensing bispectrum. In the case of S4, the lensing 
bispectrum will be detected with high statistical significance ($\agt 50\sigma$). Note that, 
for S3-wide and S3-deep, the signal-to-noise ratio is almost unchanged even if we use the quadratic 
estimator developed in Ref.~\cite{Hu:2001kj}. On the other hand, the signal-to-noise ratio with 
the quadratic estimator decreases by $\sim 20$\% for the S4 experiment. 

We check that the most dominant contribution to the signal-to-noise ratio comes from the enfolded bispectrum. 
As shown in Fig.~\ref{fig:blll}, the tree-level terms significantly contribute 
to the enfolded configuration even at $\l\sim 2000$. 
This fact indicates that terms beyond the tree level are not so significant 
unless we include the lensing signals only at $\l\alt2000$. 

\subsection{Cosmological Parameter Constraints} \label{result:cp}

\begin{table}
\bc
\caption{
Expected $1\sigma$ constraints on the beyond-$\Lambda$CDM parameters (the dark-energy 
equation-of-state $w$ and the sum of neutrino masses $\sum m_\nu$), using the lensing power spectrum (2pt), 
the lensing bispectrum (3pt), and both the lensing power spectrum and bispectrum (2pt+3pt). The prior 
information from the primary CMB anisotropies is included in all cases. The noise spectra are computed 
assuming the CMB Stage-IV. We also show the improvements by adding the lensing bispectrum to 
the lensing power spectrum. 
}
\label{Table:const} \vs{0.5}
\begin{tabular}{lcccc} \hline 
 & 2pt & 3pt & 2pt+3pt & Improvement (\%)\\ \hline 
$w$                & 0.16 & 0.21 & 0.12 & 33\% \\ \hline
$\sum m_\nu$ [meV] & 74 & 68 & 55 & 35\% \\ \hline
\end{tabular}
\ec
\end{table}

A precisely measured bispectrum would be useful to explore various issues in cosmology. As an example 
of cosmological applications, we here discuss improvement on cosmological parameter constraints if 
the lensing bispectrum is further included in the analysis. 

We compute expected constraints on cosmological parameters based on the Fisher matrix approach. 
Following TJ04, the Fisher information matrix of the lensing bispectrum is given by \cite{TJ04}
\al{
	F_{ij} = f_{\rm sky} \sum_{\l_1\leq\l_2\leq\l_3} 
		\frac{B^\kappa_{\l_1\l_2\l_3,i}B^\kappa_{\l_1\l_2\l_3,j}}{\Delta_{\l_1\l_2\l_3}\mC{C}_{\l_1}\mC{C}_{\l_2}\mC{C}_{\l_3}}
	\,, \label{Eq:fisher}
}
where $B^\kappa_{\l_1\l_2\l_3,i}$ is the derivative of the lensing bispectrum with respect to the $i$th 
cosmological parameter. We choose the maximum multipole of the summation of Eq.~\eqref{Eq:fisher} as 
$\l=2000$. In addition to the Fisher matrix of the lensing bispectrum, the Fisher matrix from 
the primary CMB anisotropies and lensing power spectrum is added in our analysis, and we ignore 
the cross covariance between the power spectrum and bispectrum of the lensing signals. We marginalize 
the six $\Lambda$CDM cosmological parameters ($\ln\Omega_{\rm b}h^2$, $\ln\Omega_{\rm m}h^2$, 
$\Omega_\Lambda$, $\ln A_{\rm s}$, $n_{\rm s}$, $\tau$), the dark-energy equation-of-state $w$, and 
the sum of the neutrino masses $\sum m_\nu$. The instrumental noise power spectrum is computed 
assuming S4. The derivatives are computed based on the symmetric difference quotient. 

Table \ref{Table:const} shows the expected $1\sigma$ constraints on $w$ and $\sum m_\nu$ using 
the lensing power spectrum (2pt), the lensing bispectrum (3pt), and both the lensing power spectrum 
and bispectrum (2pt+3pt). In all cases, we compute the Fisher matrix by adding the prior information 
matrix from the CMB temperature and E-mode polarization following e.g. Eqs.~(4.4) and (4.5) of \cite{Hannestad:2006}. 
Note that the constraints from the lensing power spectrum are consistent with other previous works 
\cite{Allison:2015,Wu:2014hta} but for a different scenario of the massive neutrinos. Comparing with 
the constraints from the lensing power spectrum (2pt), the inclusion of the lensing bispectrum (2pt+3pt) 
improves the constraint on the dark-energy equation-of-state $w$ and the sum of neutrino masses $\sum m_\nu$ 
by $33$\% and $35$\%, respectively. We find that the constraints on $w$ and $\sum m_\nu$ from 
the lensing bispectrum (3pt) are comparable to those from the lensing power spectrum (2pt). Note that 
the dark-energy density, $\Omega_\Lambda$, is also improved by $\sim 20$\%. These results indicate 
that the lensing bispectrum measured by the S4 experiment could provide additional information on 
the dark energy and the neutrino masses, and also be used for other cosmological purposes.

\section{Summary} \label{summary}

We discussed the detectability of the CMB lensing bispectrum and its cosmological application in the 
near future CMB experiments such as Advanced ACT, Simons Array, SPT3G, and S4. We found that the 
lensing bispectrum is detectable even from the near term CMB experiments. In the case of S4, the 
lensing bispectrum would be detected with high statistical significance ($\agt 50\sigma$). We then 
showed that the inclusion of the lensing bispectrum measurement further improves the constraints on 
the dark-energy parameters ($w$ and $\Omega_\Lambda$) and the sum of neutrino masses. 

We have focused on the lensing bispectrum obtained from the CMB experiments. The cross bispectrum 
between other cosmological observables such as the galaxy clustering, cosmic shear, and cosmic 
infrared background may be also detectable in ongoing and future experiments, and is important to be 
investigated. 

We have made several simple assumptions in our estimation to compute analytically. Although we assume 
the Gaussian covariance of the bispectrum to estimate the signal to noise, the non-Gaussian covariance 
may degrade the sensitivity to the lensing bispectrum, and therefore cosmology \cite{Sefusatti:2006}. 
In the Fisher matrix, we also ignored the cross covariance between the power spectrum and bispectrum 
which comes from the non-Gaussianity of the lensing signals and could degrade the cosmological constraints. 
From these respects, the expected signal to noise and constraints obtained in this paper are considered 
as their upper limits, though the non-Gaussianity of the lensing signals is expected to be much 
less significant compared to the case with the galaxy lensing shown in TJ04. The precision of 
the fitting formula \cite{Gil-Marin:2012} could also affect the resultant signal to noise ratio and 
parameter constraints. The fitting formula should be therefore tested against a wide range of 
cosmological parameters for a robust forecast and a realistic cosmological analysis. Even in 
the absence of the nonlinear gravitational potential, the post-Born correction generates a bispectrum 
in the observed lensing signals, and could bias in estimating the nonlinear growth \cite{Pratten:2016}. 
The validity of our assumptions is worth investigating, and will be addressed in our future work. 
Albeit simple, our results definitely show that the nonlinear evolution will be no longer negligible 
in ongoing and near future CMB experiments, and will have fruitful cosmological information 
comparable to that from the lensing power spectrum. 

At the time of writing this paper, Boehm {\it et al}.~\cite{Boehm:2016} explore the bias in the 
lensing reconstruction induced by the lensing bispectrum, finding that there is an additional 
non-negligible bias in estimating the lensing power spectrum in the case of S4. Their result also 
implies that, in the era of S4, the effect of the nonlinear growth should be properly taken into 
account even in the CMB lensing analysis. 

\ifdefined\compile
\begin{acknowledgments}
We thank Vanessa Boehm and Blake Sherwin for cross-checking part of our results, and we are also 
grateful to Marcel Schmittfull and Atsushi Taruya for enlightening comments. This work is supported 
in part by JSPS Postdoctoral Fellowships for Research Abroad No.~26-142. 
\end{acknowledgments}

\bibliographystyle{apsrev}
\bibliography{cite}
\fi

\end{document}